\documentclass[twoside]{article}
\usepackage[a4paper]{geometry}
\usepackage[latin1]{inputenc} 
\usepackage[T1]{fontenc} 
\usepackage{RR}
\usepackage{hyperref}

\usepackage{amsmath,epsfig,graphics,graphicx}
\usepackage{amsfonts,amssymb}
\usepackage{color}
\usepackage{mathrsfs}
\usepackage{enumitem}
\usepackage{array}
\usepackage{float}
\usepackage{alltt}
\usepackage{verbatim}

\newcommand{\tet}[1]{\texttt{#1}}
\newcommand{\tit}[1]{\textit{#1}}
\newcommand{\model}{\textbf{\textsc{model}}}

\newcommand{\deriv}[2]{\frac{\partial{#1}}{\partial{#2}}}

\newcommand{\mytilde}{\raise.17ex\hbox{$\scriptstyle\sim$}}

\definecolor{darkgreen}{RGB}{0,142,128}
\definecolor{darkgray}{RGB}{102,102,102}
\definecolor{lightgray}{RGB}{240,240,240}

\RRNo{8134}
\RRdate{November 2012}
\RRauthor{
G. Latu\thanks[sfn]{CEA Cadarache, IRFM bat. 513, F-13108 Saint-Paul-lez-Durance}%
\and M. Becoulet\thanksref{sfn}
\and G. Dif-pradalier\thanksref{sfn} 
\and V. Grandgirard\thanksref{sfn} \and \\
M. Hoelzl\thanks{Max-Planck-Institute for Plasma physics, Boltzmannstr. 2, D-85748 Garching}
\and G. Huysmans\thanks{ITER Organisation, Route de Vinon sur Verdon, F-13115 St-Paul-lez-Durance}
\and X. Lacoste\thanks[inr]{INRIA, 351 cours de la Liberation - F-33405 Talence}
\and E. Nardon\thanksref{sfn}
\and F. Orain\thanksref{sfn} \and\\
C. Passeron\thanksref{sfn} 
\and P. Ramet\thanksref{inr}
\and A. Ratnani\thanksref{sfn} 
}
\authorhead{Latu \& al}
\RRtitle{Tests de non-regression pour le code JOREK}
\RRetitle{Non regression testing for the JOREK code}
\titlehead{NRT for JOREK}

\RRresume{ Les tests de non regression (l'acronyme anglais est NRT)
  ont pour objet de v\'{e}rifier si les modifications apport\'{e}es
  \`{a} un logiciel conduisent, ou non, \`{a} des comportements
  corrects ou incorrects.  Ayant caract\'{e}ris\'{e} et
  r\'{e}f\'{e}renc\'{e} des comportements corrects li\'{e}s \`{a} des
  sc\'{e}narii d'ex\'{e}cution pr\'{e}cis, ces tests permettent
  d'identifier durant le processus de d\'{e}veloppement une
  \'{e}ventuelle r\'{e}gression, un bug. L'am\'{e}lioration et
  l'optimisation d'un code parall\`{e}le est une t\^{a}che
  consommatrice de temps et parfois difficile. Cela est d'autant plus
  vrai lorsque diff\'{e}rents acteurs interagissent \'{e}troitement,
  dans notre cas~: des physiciens, des math\'{e}maticiens, des
  informaticiens. Le code JOREK traite d'instabilit\'{e}s li\'{e}es
  \`{a} la Magn\'{e}tohydrodynamique (MHD) dans des plamas de
  Tokamak. Ce papier d\'{e}crit la proc\'{e}dure de NRT qui a
  \'{e}t\'{e} mise en place dans ce code de
  simulation. L'automatisation des NRT est un point essentiel pour
  conserver, dans la dur\'{e}e, un code sain dans un d\'{e}p\^{o}t de
  sources. }

\RRabstract{ Non Regression Testing (NRT) aims to check if software
  modifications result in undesired behaviour. Suppose the behaviour of
  the application previously known, this kind of test makes it possible to
  identify an eventual regression, a bug. Improving and tuning a
  parallel code can be a time-consuming and difficult task, especially
  whenever people from different scientific fields interact
  closely. The JOREK code aims at investing Magnetohydrodynamic (MHD)
  instabilities in a Tokamak plasma. This paper describes the NRT
  procedure that has been tuned for this simulation
  code. Automation of the NRT is one keypoint to keeping the code
  healthy in a source code repository.  
}

\RRmotcle{MHD non-lin\'{e}aire, plasma de Tokamak, tests de non-regression}
\RRkeyword{nonlinear MHD, Tokamak plasma, Non Regression Testing}
\RRprojet{HIEPACS}
\RCBordeaux 

\begin{document}
\makeRR   

\section{Introduction}

This paper deals with a Non Regression Testing (NRT) tool for a
code dedicated to the simulation of MHD instabilities relevant to
magnetically confined fusion.

Magnetohydrodynamic (MHD) stability and the avoidance of plasma
disruption - rapid loss of plasma energy and abrupt termination of the
plasma current caused by global MHD instability growth - are key
considerations to the attainment of burning plasma conditions in ITER
(a large fusion reactor 
\url{www.iter.org}\,).  Realization of adequate MHD stability and
careful control of the plasma operation will be critical ITER
operation issues: MHD instabilities can damage components of tokamak
walls.

Numerical simulations play an important role in the investigation of
the non-linear behaviour of these instabilities and can help for
interpretating experimental observations. In the framework of
non-linear MHD codes, targeting realistic simulation requires: to
model complex geometry, to handle a large gap between the different
time scales relevant to plasmas, to address full 3D
simulation. Computational time needed to run 3D MHD code named JOREK
necessitates parallel computing in order to get reduced restitution time
for the user\,\cite{czarny08,hoelzl12}.

We describe here procedures and solutions to overcome a lack of NRT
and benchmarking tools for JOREK users. The aims of such improvements
are twofold: first, to keep the main JOREK trunk healthy in the
version control system, second, to have a way to compare execution times
and results of runs launched on different supercomputers.

In this document, we will often refer to the SVN repository where
JOREK code is stored. This repository contains also all the materials
and scripts that are described herein:\\[-3mm]
\begin{center}
\tet{scm.gforge.inria.fr/svnroot/aster}
\end{center}

This work was made possible by a ANR grant and the accesses to several
parallel machines.  The authors acknowledge the support of the French
Agence Nationale de la Recherche (ANR) under reference
ANR-11-MONU-0002 (ANEMOS project). Computations have been performed at
the M\'{e}socentre d'Aix-Marseille Universit\'{e} (France), on PLAFRIM
Bordeaux (France), on IFERC Rokasho (Japan).

\section{NRT}

To begin with, we will sketch up the JOREK environment in order to
explain the choices made for the Non Regression Testing. Then, the method
used to achieve NRT is explained.

\subsection{Numerical components}
\label{num_comp}
\paragraph{Spatial discretization}
JOREK is a MHD three dimensional fluid code that takes into account
realistic tokamak geometry. High spatial resolution in the poloidal plane
is needed to resolve the MHD instabilities at high Reynolds and/or
Lundquist numbers.  Bezier patches (2D cubic Bezier finite elements)
are used to dicretize variables in this plane. Hence, several physical
variables and their derivatives have a continuous representation over
a poloidal cross-section. The periodic toroidal direction is treated
via a sine/cosine expansion.

\paragraph{Set of equations}
Some of the variables modelled in JOREK code are: the poloidal flux
$\Psi$, the electric potential $u$, the toroidal current density $j$,
the toroidal vorticity $\omega$, the density $\rho$, the temperature
$T$, and the velocity $v_{parallel}$ along magnetic field
lines. Depending on the model choosen (hereafter denoted \model{}
which is a simulation parameter), the number of variables and the
number of equations on them are setup. At every time-step, this set of
reduced MHD equations is solved in weak form as a large sparse
implicit system. The fully implicit method leads to very large sparse
matrices. There are some benefits to this approach: there is no
\tit{a priori} limit on the time step, the numerical core adapts
easily on the physics modelled (compared to semi-implicit methods that
rely on additional hypothesis). There are also some disadvantages:
high computational costs and high memory consumption for the parallel
direct sparse solver (PASTIX or others).

\paragraph{Time integration scheme}
The temporal discretization is performed by a fully implicit
second-order linearized Crank-Nicholson scheme. Two operators $C$ and
$D$ are matrices that describe the set of MHD equations in weak form:
\begin{eqnarray}
&&\deriv{C(\vec{u})}{t}=D(\vec{u}),\ \ \ {\rm discretized\ by\  }
\left(\deriv{C(\vec{u_n})}{u}-\frac{1}{2}\delta_t\,\deriv{D(\vec{u_n})}{u}\right)\delta\vec{u}= D(\vec{u_n})\,\delta{}t\\
&&{\rm Let\ us\ denote\ \ } A=\deriv{C(\vec{u_n})}{u}-\frac{1}{2}\delta_t\,\deriv{D(\vec{u_n})}{u},\ \ b=D(\vec{u_n})\,\delta{}t,\ \ {\rm then}\\
&&A\,\delta{}\vec{u} = b \label{ax_b}
\end{eqnarray}
Where $\vec{u_n}$ is the set of variables at time step $n$, and
$\delta{}\vec{u}=\vec{u_{n+1}}-\vec{u_{n}}$ is the unknown. In the
following, we will refer to $A$ as the sparse matrix that should be
solved, and $b$ the right hand side of the problem.

\paragraph{Equilibrium}
Each JOREK simulation begins with the solving of the static magnetic
equilibrium equation (so-called Grad-Shafranov equation) in the 2
dimensions of the cross-section plane. A keypoint is the ability to
handle magnetic equilibria which include an X-point. High accuracy is
needed to have a correct representation of this equilibrium and avoid
spurious instabilities whenever the whole simulation 3D+t is launched. 


JOREK is able to build a Bezier finite element grid aligned with the
flux surfaces both inside and outside the separatrix (\tit{i.e.} the
flux surface containing the X-point). This strategy allows one to
improve the accuracy of the equilibrium representation.  The flux
surfaces are represented by sets of 1D Bezier curves determined from
the numerical solution of the equilibrium. The Grad-Shafranov solver
is based on a Picard's iteration scheme.

After the Grad-Shafranov solving step, a supplementary phase is
required: the time-evolution equations are solved only for the
$n\!=\!0$ mode (the first toroidal harmonic, \tit{i.e.} purely
axisymmetric) over a short duration. First, very small time-steps are
taken, then they are gradually increased. This process allows the
plasma equilibrium flows to establish safely in simulations involving
a X-point.

\paragraph{Sparse solver \& preconditionning}

A direct parallel sparse matrix solver (PASTIX or others) is used to
solve the large linear systems inside JOREK. In order to minimise the
memory requirements of the fully implicit solver and to access larger
domain sizes, a preconditioner accompanied with a GMRES iterative solver
have been included a few years ago.  Preconditioning
transforms the original linear system $A\,x\!=\!b$ into an equivalent one
which is easier to solve by an iterative technique. A good
preconditioner $P$ is an approximation for $A$ which can be
efficiently inverted, chosen in a way that using $P^{-1}\,A$ or
$A\,P^{-1}$ instead of $A$ leads to a better convergence
behaviour. Usually, GMRES iterative solver is applied in
collaboration with a preconditioner. The preconditioner typically
incorporates information about the specific problem under
consideration.

The JOREK \tit{physics-based} preconditioner has been constructed by
using the diagonal block for each of the \tet{n\_tor} Fourier modes in
the toroidal direction of the matrix $A$ presented earlier in
Eq.~\ref{ax_b}. The preconditioner represents the linear part of each
harmonic but neglects the interaction between harmonics (similar to a
block-Jacobi preconditioning on a reordered matrix). So, we set many
coefficients of the original matrix $A$ to zero, in order to get a
block-diagonal matrix with $m$ independent submatrices on the
diagonal. The preconditioner $P$ consists in the composition of $m$
independent linear systems $\left(P^\star_i\right)_{i\in[1,m]}$, with
$m=\frac{n\_tor+1}{2}$. Practically, the set of processors are split
in $m$ independent MPI communicators, each of them treats only one
single linear system $P^\star_i$ with a sparse direct solver. This
preconditioned parallel approach avoids large costs in terms of memory
consumption compared to the first approach that considers the whole
linear system to solve (it saves the memory needed by the sparse
solver to store the decomposition of big matrix $A$ - \tit{e.g.} $L$,
$U$ factors). Nevertheless, the whole linear system $A$ has to be
built (in parallel) in order to perform the matrix-vector
multiplication needed by the GMRES. But, the cost in memory and in
computation is far less than invoking the parallel sparse solver on
the large linear system $A$.

This strategy improves the scalability ($m$ inpendent systems), and
the parallelisation performance of the code. The bad point is that in
some specific circumstances, the iterative scheme may not converge.

\subsection{Parameters}
From the user perspective, one can distinguish three sets of
parameters in the JOREK code:
\begin{itemize}
\item\textbf{Category A - Hardcoded.} A first set of parameters is fixed inside the
  source code. For example: the selected physics model (\model{}
  parameter), the number of harmonics which defines the discretization
  along toroidal dimension (\tet{n\_tor}), and some others
  (\tet{n\_period}, \tet{n\_plane}, \tet{n\_vertex\_max},
  \tet{n\_nodes\_max}, \tet{n\_elements\_max}, \tet{n\_boundary\_max},
  \tet{n\_pieces\_max}, \tet{gmres\_max\_iter}, \tet{iter\_precon},
  $\ldots$).
\item\textbf{Category B - Input file.} A second set of parameters are given through
  the standard input to JOREK at launch time. Some examples of such
  input files can be found in \tet{jorek2/trunk/namelist} directory in the JOREK
  repository.
\item\textbf{Category C - Environment.} A third set of parameters consists in the
  parallel environment at launch time. For example, the following data
  constrain the execution time and impact a little bit the
  numerical results: the number of MPI processes, the number of shared
  memory nodes, the number of OpenMP threads, the set of libraries
  used (MPI, sparse solver library).
\end{itemize}

Several kinds of simulations can be undertaken with this tool. We will
mainly address in this document few simulations that are parametrized
by the following inputs (mainly of category~B - input file):
\begin{itemize}
\item \textbf{Geometry} Three main configurations are available
  concerning the geometry in the poloidal plane: circular
  cross-section (parameter \tet{xpoint=.f.}), single X-point
  (parameter \tet{xpoint=.t.}), two X-points (not detailed
  here). Other fine parameters permit to fit the geometry to a
  realistic configuration.
\item \textbf{Spatial domain size} Some parameters set up the
  computational domain. The discretization highly impacts the memory
  and computational costs. Poloidal domain is sized by \tet{n\_flux}
  the number of flux surfaces (radial coordinate), \tet{n\_tht} the
  number of points in the angular direction (equivalent to
  $\theta$). In the toroidal direction, \tet{n\_tor-1} sizes the number
  of sine/cosine components that will be computed. The \tet{n\_tor}
  parameter is in the \mbox{category A - hardcoded}. JOREK code is able to handle
  the increase of \tet{n\_tor} parameter between a checkpoint and a
  restart.
\item \textbf{Time axis} Depending on the simulation kind and the
  evolution of the simulation, the time step duration can be
  adjusted. The timestep \tet{tstep} is typically fixed from a fraction
  of an Alfven time to thousands of Alfven times. The number of time steps
  in a single run is set by \tet{nstep}. There is another way to
  specify both number of time steps and durations through the parameters
  \tet{tstep\_n} and \tet{nstep\_n} which are vectors. It helps the user
  in order to define complex scenarii with multiple timesteps inside a
  single run.
\end{itemize}

\subsection{Scenarii definition}

Edge localized modes (ELMs) are intermittent relaxations of the edge
transport barrier. During these events, particles and energy are
ejected from the plasma edge into the surrounding low density
envelope, the SOL (scrape-off layer). The associated transient power
loads create problems during the operation of tokamak reactors. MHD
codes, such as JOREK, can describe the ELM evolution.

One of the relevant instabilities for the ELMs are the MHD
peeling-ballooning modes, but other kinds of instabilities or physics
phenomena can be studied with JOREK\,\cite{huysmans07,nardon07}
\mbox{(kink modes, tearing modes ...)}. A typical simulation begins
from an axisymmetric equilibrium ($n\!=\!0$) on top of which a small
$n\ne{}0$ perturbation is initialised. First, an initial linear phase
of exponential growth shows a rapid energy increase (when a mode is
unstable), starting from a very low level of energy (a little bit
noisy). Kinetic and magnetic energies can have quite large growth
rates during the linear phase. For some identified cases, the theoretical
growth rates that should be obtained are known. After a while, some
Fourier modes saturate non-linearly (growth rates are droping).

In order to track a simulation and analyse its time evolution, we will
look at the dynamics of kinetic and magnetic energies. Somehow, they
constitute a valid signature of a given simulation scenario. The
physics phenomena strongly constrain these energies. Other criteria
could also be taken into account in the future, in order to track fine
differences between two runs that reproduce the same scenario (for
example spatially localized phenomena).

A typical scenario execution is decomposed in the following
way. First, the equilibrium is computed separately within a single
JOREK execution. Second, a set of \tet{tstep\_n} and \tet{nstep\_n}
vectors defines a \tit{subsequence} of time steps that will be used in
a second JOREK execution.  A complete scenario is composed from one
\tit{subsequence} to several such \tit{subsequences}.  Let us notice
that between two successive JOREK executions, the parameter
\tet{n\_tor} can be modified in order to increase/decrease toroidal
resolution.  A set of scenarii definitions have been encapsulated in
\tet{jorek2/trunk/util/launch\_tests.sh}. For each scenario, the
following data are fixed: the \tit{input} parameter file, the set of
vectors \tet{tstep\_n} and \tet{nstep\_n}, the times where
\tet{n\_tor} is changed, and the series of JOREK executions (2 to 9
successive runs).

\subsection{Metric}

In order to build \tit{Non Regression Testing} for a whole simulation
code\footnote{Non Regression Testing involve an entire code whereas
  unitary tests involve subroutines or subparts of the code.},
one mainly needs two basic tools: 1) a methodology to replay a given
scenario for the code, 2) a way to measure the differences between two
runs of the same scenario with respect to test selection criteria. The
second point involves the definition of a \tit{Metric} that captures
numerically the distance beween two scenario execution
traces. Ideally, the \tit{Metric} should also allow users to identify
problems, help them in the process of improving the code, and provide
a way to detect bugs quickly.

Reproducibility is the degree of agreement between measurements
conducted on replicate scenario. While repeatability of scientific
experiments is desirable, this is not always an easy task, even in the
field of numerical simulations. Typically in the JOREK code, there are
at least four reasons that limit someone to reproduce the same
results upon demand:
\begin{itemize}
\item OpenMP introduces a dependance on threads scheduling inside
  processes that may alter a bit numerical results from one run to
  another,
\item Global summation with MPI of distributed arrays is most
  susceptible to nonreproducible rounding errors (\tit{e.g.} addition in
  MPI\_Reduce is not strictly commutative),
\item The iterative solver used for the large linear systems depends
  on a threshold to stop the iteration loop, this nonlinearity is able to
  amplify numerical noise,
\item In many simulations, the starting point of the linear phase is
  dominated by numerical noise because system state is not far from a perfect
  equilibrium; then, the very beginning of a simulation run is expected
  to be very difficult to reproduce. Nevertheless, the rest of the
  simulation is dominated by a strong signal with much
  less numerical noise and this second part is reproducible.
\end{itemize}

So, given that the application is nondeterministic, the difficulty is to
discriminate differences between two runs that are fully acceptable
(coming from nonreproducibility), from differences that are likely to
originate from a bug.

The metric we have chosen is the following: \tit{a time series of the
  growth rate of magnetic and kinetic Fourier modes on a specific time
  interval}. The number of Fourier modes and the time interval are
test-case dependent. The time interval will be fixed such that the very
beginning of the simulation is excluded. We will also not keep the
long term evolution of the non-linear saturation because small noise can
be amplified and therefore comparison is notably biaised. Finally,
only a time interval in the linear phase is kept for this
metric. Practically, this means that we have for each reference scenario: a
time interval $[t_{start},t_{end}]$ and a threshold $thr$, a time
series $S$ of the growth rates of $M$ Fourier modes. Let us denote
$s_A^{t,m}$ the growth rates of Fourier mode $m\in[1,M]$ at time step
$t\in[t_{start},t_{end}]$ of simulation $S_A$.  To compare two
different runs $A$ and $B$ reproducing the same scenario, the metric
consists in comparing $S_A$ and $S_B$ with the following criteria:

$$\forall t,\ \forall m,\ \ \ \ \ \frac{2\,\lvert{}s_A^{t,m}-s_B^{t,m}\rvert{}}{\lvert{}s_A^{t,m}\rvert+\lvert{}s_B^{t,m}\rvert{}} < thr$$

The difference between two growth rates coming from two runs A and B
should be below a given threshold percentage fixed by $thr$. If it is not the case, we will consider that $A$ and $B$ are different, \tit{i.e.} one run has a wrong behaviour.

\newpage
\subsection{Parallel job launch}

For the sake of simplicity, a complete scenario is executed in only
one single parallel job. If it was not the case, one would have to
chain multiple jobs which requires some specific parameters to give to
the batch scheduler of the parallel machine. In order to automate the
NRT as much as possible, the compilation and execution are done within
the same parallel job (three models are available: 199, 302, 303).
There are two possibilities to launch a JOREK reference scenario:
\begin{enumerate}
\item Launch a parallel job over 4 SMP nodes (one MPI process per
  node) and inside this job compile all executables and then run the
  scenario with the following commands (it requires that you have
  configured your Makefile.inc already):\\[1mm] 
\scalebox{.80}{
    \colorbox{lightgray}{
\parbox{1.20\linewidth}{
\begin{alltt}
\$ jorek2/trunk/util/compile\_test.sh <MODEL> \# compile JOREK for some <MODEL>\\
\$ export PRERUN='export OMP\_NUM\_THREADS=8' \ \# command to launch before mpirun\\
\$ export MPIRUN='mpirun -np 4'\ \ \ \ \ \ \ \ \ \ \ \ \ \  \# mpirun command\\
\$ export BASEDIR='/scratchdir/username' \ \ \ \   \# root where the new dir will be created\\
\$ jorek2/trunk/util/launch\_test.sh <DIRECTORY PREFIX><MODEL> \# run the scenario
\end{alltt}}}}
Let us remark that on some machines (\tit{e.g.} Helios-IFERC machine), 
the compilation process (\tit{i.e} \tet{compile\_test.sh}) can not 
be done on compute nodes. Therefore the compilation should be done 
on login nodes, before the job launch.
\item If \tet{subjorek} is configured for the target parallel machine,
  you can alternatively use the following command:\\[1mm]
\scalebox{.80}{
    \colorbox{lightgray}{
\parbox{1.20\linewidth}{
\begin{alltt}
jorek2/trunk/run\_jorek/nrt\_run.sh all <DIRECTORY PREFIX><MODEL>
\end{alltt}}}}
\end{enumerate}

The reference scenario will be executed and will
generate a directory named \tet{<DIRECTORY PREFIX><MODEL>} that
contains all outputs of a standard JOREK run. In this directory, the
main data that we will used to compare numerical results to another run is
stored in the file \tet{macroscopic\_vars.dat}.

Some examples of batch scripts to run a reference scenario are given
in \tet{benchmark/batch} directory. 
\subsection{Comparing against reference scenario}

To compare a run against a reference scenario, you have to download
some files from the SVN repository with a command such as\\
\hspace*{1cm}\tet{svn co
  svn+ssh://<LOGIN>@scm.gforge.inria.fr/svnroot/aster/benchmark}\\ A
set of reference scenarii is located in \tet{benchmark/model*}
directories.  The data used to compare with the metric a JOREK run
against a reference scenario is encapsulated into the script file
\tet{jorek2/trunk/util/nrt\_compare.sh}. This script requires the
access to a tiny tool named \tet{numdiff} that is easy
to compile and install on a LINUX/UNIX system
(\url{http://ftp.igh.cnrs.fr/pub/nongnu/numdiff/}).

For example, let suppose that you have reproduced the reference
scenario of model \tet{199} with the command:\\[1mm]
\scalebox{.80}{ \colorbox{lightgray}{
\parbox{1.05\linewidth}{
\begin{alltt}
\$ \tet{jorek2/trunk/util/launch\_test.sh\ new199}
\end{alltt}
}}} \\[1mm] 
After the job completion, the directory \tet{new199}
contains the results of the simulation you have launched. To compare
your data with the reference case stored into
\mbox{\tet{aster/benchmark/model199/helios\_a}}, you just have to do
this:\\[1mm] 
\scalebox{.80}{ \colorbox{lightgray}{
\parbox{1.05\linewidth}{
\begin{alltt}
\$ nrt\_compare.sh 199 new199 aster/benchmark/model199/helios\_a\\
OK  (nb lines compared: 53/53)\\
 \\
\# gnuplot commands to look at growth rates that have been compared :\\
    set key autotitle columnhead;\\
    set auto; plot 'f1' u 1:2 ls 1, 'f2' u 1:2 ls 4; pause -1\\
    set auto; plot 'f1' u 1:3 ls 3, 'f2' u 1:3 ls 6
\end{alltt}
}}}\\[1mm] 
The comparison is successful and 53 time steps over 53 have
been compared and verified\footnote{It is also possible to perform a
  \tet{nrt\_compare.sh} during a simulation and the check will be
  realized on available data in the \tet{macroscopic\_vars.dat}
  file.}.  To look at the growth rates of kinetic and magnetic
energies that have been compared, you can use the gnuplot commands
that are mentionned. The data inside \tet{f1} and \tet{f2} files are
extracted from growth rates of the file \tet{macroscopic\_vars.dat}
that is generated by JOREK.

\begin{figure}[H]
\centering
\begin{minipage}[h]{.53\textwidth}
\includegraphics[trim=20mm 20mm 60mm 55mm,clip,width=\textwidth]{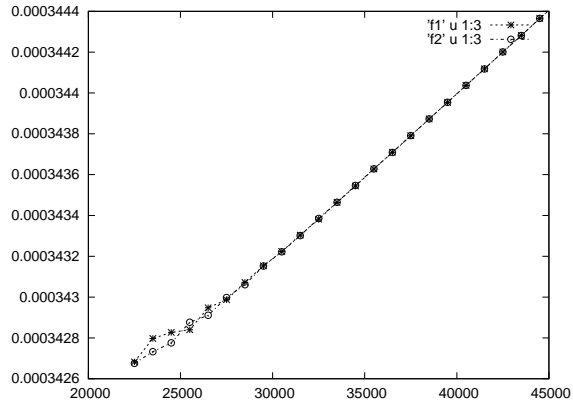}
\end{minipage}
\caption{Example of gnuplot output generated by plotting f1 and f2
  files with time evolution (abscissa) of growth rates (ordinate)}
\label{f1ex}
\end{figure}

On the Fig. \ref{f1ex}, the \tet{f1} curve corresponds to the growth rate
of one Fourier mode (magnetic energy $n\!=\!3$) for simulation
\tet{model199/helios\_a}, while \tet{f2} corresponds to the same mode
on another machine \tet{new199} (\tet{= model199/rheticus\_a}). One can
see that even the beginning differs slightly (numerical noise is
perceptible), the two curves coincide almost exactly above 30\,000
Alfven times. The \tet{nrt\_compare.sh} script checks between the two
simulations if growth rates do not differ much more than 1 percent and
says \tet{OK} in this case.\\

A good methology using NRT is the following: whenever the user wants
to perform a \mbox{\tet{svn commit}}, he or she should launch all
reference scenarii. If the modification is not supposed to alter the
results, \tet{nrt\_compare.sh} will say \tet{OK} for each
scenario. The database for Non Regression Testing is located in
\tet{benchmark/model*}, the user can equally refer to all cases stored
into \tet{benchmark/model*} directories as third parameter of the
\tet{nrt\_compare.sh} call. Once the user has obtained a \tet{OK} for
all scenarii, he or she can commit the changes in the SVN repository.

\section{Benchmarks}

The need for NRT has permitted to define a set of scenarii which are
representative of some classical user cases of JOREK (yet, not
exhaustive but the list is expected to be extended by users). These
scenarii can also be used to benchmark machines and to compare execution
times in different parallel configurations (number of MPI processes,
threads), on several parallel machines. This information helps the
user to determine if the performance on a new machine is far away
from other machines or not, and to evaluate
the parallel performance.

\subsection{Timer measurements}

A script \tet{timing\_bench.sh} is located in the \tet{benchmark}
directory. It facilitates the extraction of timers stored in JOREK log
files. For example to look at the elapsed time taken by each JOREK temporal
iteration on the two machines \tet{fourmi} and \tet{rheticus} in the
\tet{out\_loop1} log file, one can execute the command:\\[1mm]
\scalebox{.80}{ \colorbox{lightgray}{
\parbox{1.05\linewidth}{
\begin{alltt}
\$ cd benchmark\\
\$ ./timing\_bench.sh ITER out\_loop1 grep3  model199/fourmi\_a model199/rheticus\_a\\
== model199/fourmi\_a ( openmpi )\\
  0              \# Elapsed time ITERATION :          22.4025180\\
  0              \# Elapsed time ITERATION :          10.5183380\\
  0              \# Elapsed time ITERATION :          11.3435830\\
== model199/rheticus\_a ( mvapich2 )\\
  0              \# Elapsed time ITERATION :          15.2480000\\
  0              \# Elapsed time ITERATION :           7.2230000\\
  0              \# Elapsed time ITERATION :           7.5020000
\end{alltt}}}}\\[1mm]
Several parameters has to be given to this script: 1) the keyword that
will be used to perform the grep in the log file (\tit{e.g.}
\tet{ITER} to get the time taken by one temporal iteration), 2) the
name of the JOREK log file (\tit{e.g.}  \tet{out\_loop1}), 3)
optionally \tet{grep} followed by a number specifying the number of
lines to keep (\tit{e.g.} \tet{grep3} to keep the 3 first lines), 4) a
list of directories where the script will lookup the log files to
process (\tit{e.g.} \tet{model199/fourmi\_a model199/rheticus\_a} to
look into theses two directories).\\ 
The outputted numbers show that
the \tet{rheticus} machine performs a single JOREK iteration a little bit
faster than the \tet{fourmi} machine. The two machines that have been used
for the runs are described in \tet{README.txt} files.\\[3mm]
\begin{minipage}[h]{.5\linewidth}
\scalebox{.70}{\colorbox{lightgray}{
\parbox{1.30\linewidth}{
\begin{alltt}
\$ cat model199/fourmi\_a/README.txt                 \\
Machine    : PLAFRIM/fourmi                         \\
Location   : Bordeaux - France                      \\
Processor  : Intel(R) Xeon(R) CPU X5550  @ 2.67GHz  \\
Cores/node : 8                                      \\
Author     : Latu                                   \\
Comment    : openmpi                                \\
Jorek fortran compiler : ifort 11.1                 \\
Jorek SVN revision     : 666                        \\
Jorek specific option  : -                          \\
Pastix compiler        : icc 11.1                   \\
Pastix SVN revision    : 3564                       \\
Pastix specific option : -                          \\
MPI library            : openmpi/1.4.4              \\
BLAS library           : GotoBLAS2                  \\
Network                : Infiniband QDR : 40Gb/s    \\
Other info             : -                          \\
\end{alltt}}}}\\[2mm]
\end{minipage}~\begin{minipage}[h]{.5\linewidth}
\scalebox{.70}{\colorbox{lightgray}{
\parbox{1.30\linewidth}{
\begin{alltt}
\$ cat model199/rheticus\_a/README.txt                \\
Machine    : Rheticus                                 \\
Location   : Marseille - France                       \\
Processor  : Intel(R) Xeon(R) CPU X5675  @ 3.07GHz    \\
Cores/node : 12                                       \\
Author     : Latu                                     \\
Comment    : mvapich2                                 \\
Jorek fortran compiler : gfortran 4.4.6               \\
Jorek SVN revision     : 666                          \\
Jorek specific option  : -                            \\
Pastix compiler        : icc 12.1.0                   \\
Pastix SVN revision    : 3564                         \\
Pastix specific option : -                            \\
MPI library            : mvapich2-1.7                 \\
BLAS library           : atlas                        \\
Network                : Infiniband                   \\
Other info             : -                            \\
\end{alltt}}}}\\[2mm]
\end{minipage}\\[2mm]
At present day, a reduced set of timers are accessible. Here is a
list of these timers:\\[-6mm]
\begin{itemize}
\itemsep=0pt\topsep=0pt\partopsep=0pt%
\parskip=0pt\parsep=0pt%
\item \textbf{construct\_matrix}: time elapsed to compute all matrix
  coefficients of the global system $A$ and to build the right hand side
  vector $b$ (see Eq. \ref{ax_b}). A domain decomposition is used to parallelize this step,
  based on distribution of the Bezier finite elements among MPI processes. A
  communication step is also performed to form the RHS $b$.
\item \textbf{distribute}: time to distribute the coefficients of the
  preconditioner matrix $P^\star_i$ (copied from distributed $A$) to all master processes,
  but also to send the local RHS associated to each local
  $P^\star_i$. There is one master process per harmonic (\tit{i.e.}
  $\frac{n\_tor+1}{2}$) that receives $P^\star_i$ and its local RHS.
\item \textbf{coicsr}: time to convert the preconditioner matrix $P^\star_i$
  on each master from $(i,j,value)$ format to CSR (standard compressed
  sparse format). This transform is generally needed to give $P^\star_i$ as input
  to the sparse parallel solver (actually Pastix or MUMPS or WSMP).
\item \textbf{analysis}: first phase of the sparse solver to solve
  linear systems for matrices $P^\star_i$. It comprises an ordering
  step that reorders the rows and columns, an analysis step or
  symbolic factorization that determines the nonzero structures of
  matrix and creates internally suitable data structures.
\item \textbf{facto}: second phase of the sparse solver, the factorisation computes
  the $L$ and $U$ matrices (or another decomposition).
\item \textbf{first solve}: time to solve the first set of linear
  systems in a single JOREK temporal step (matrices $P^\star_i$), that
  includes potentially the analysis, the factorization and the solve
  step (forward and back substitution algorithms). But often, analysis
  or factorisation can be skipped.
\item \textbf{gmres/solve}: time needed by GMRES iterative process
  that solves many times the linear system until a convergence is
  achieved or a maximum iteration count is reached. Because analysis
  and factorization are not done again and again, this gmres/solve
  step is mainly $m$ successive solve steps using $P^\star_i$ matrix.
\item \textbf{ITER}: time for one temporal iteration of JOREK. It can
  be viewed as a partial sum of a subset of the previous timing
  counters.
\end{itemize}

Depending on the configuration, the compilation and the deployment of
JOREK on a particular parallel machine, the relative weight of these
timers over the total time can be quite different. For example, some parts
scale along with the number of cores used (\tit{e.g.}
\tet{construct\_matrix}), others do not scale well (\tit{e.g.}
\tet{coicsr}).\\

If you want to store the data obtained from a reference run in the SVN
repository (your run is saved into \mbox{\mytilde{}/MYDIR\_302} for
example), you can use the following
procedure:\\ \scalebox{.80}{\colorbox{lightgray}{
\parbox{1.05\linewidth}{
\begin{alltt}
\$ cd benchmark\\
\$ ./cp\_bench.sh \mytilde{}/MYDIR\_302 model302/MYMACHINE\_a
\end{alltt}
}}}\\[1mm] 

This will create a directory at the
\tet{benchmark/model302/MYMACHINE\_a} location and also copy from your
directory \tet{\~{}/MYDIR\_302} only a reduced set of files that
should be conserved (the \tet{macroscopic\_vars.dat} file, the
standard ouput files of JOREK runs). There will be a \tet{README.txt}
file created in the new directory that you can fill in order to give
some information on the machine you used.


\subsection{Comparing timers}

A set of benchmarks of JOREK on several machines and with different
settings are presented in this section. It can give some clues to the
JOREK user in order to choose a convenient set of parameters during
the configuration/compilation of the code on a new machine.

\subsubsection{Impact of the MPI library}

On Helios machine (IFERC, Rokasho-Japan), available versions of MPI
can not use MPI\_THREAD\_MULTIPLE safely (up to now - september
2012). Then the \tit{funneled} version of PASTIX has to be used
(MPI\_THREAD\_FUNNELED mode of MPI).\\ Two MPI libraries have been
tested in order to determine which one should be preferred: \tet{intelmpi}
(version 4.0.3) or \tet{bullxmpi} (version 1.1.14.3).\\[2mm]
\scalebox{.80}{\colorbox{lightgray}{
\parbox{1.05\linewidth}{
\begin{alltt}
\$ ./timing\_bench.sh gmres/solve out\_loop5 grep2 model302/helios\_?  \\
== model302/helios\_a ( bullxmpi + FUNNELED )                         \\
  0            \# Elapsed time gmres/solve :           3.8031330      \\
  0            \# Elapsed time gmres/solve :           3.4623600      \\
== model302/helios\_b ( intelmpi + FUNNELED )                         \\
  0            \# Elapsed time gmres/solve :           3.4759830      \\
  0            \# Elapsed time gmres/solve :           3.3056950      \\
\$ ./timing\_bench.sh facto out\_loop5 grep1 model302/helios\_?   \\
== model302/helios\_a ( bullxmpi + FUNNELED )                     \\
  0                \#\# Elapsed time, facto :          36.4806480  \\
== model302/helios\_b ( intelmpi + FUNNELED )                     \\
  0                \#\# Elapsed time, facto :         143.5415650  \\
\$ ./timing\_bench.sh ITER out\_loop5 grep5 model302/helios\_?     \\
== model302/helios\_a ( bullxmpi + FUNNELED )                     \\
  0              \# Elapsed time ITERATION :         151.7923330   \\
  0              \# Elapsed time ITERATION :          62.5559670   \\
  0              \# Elapsed time ITERATION :           9.5537380   \\
  0              \# Elapsed time ITERATION :          10.3767240   \\
  0              \# Elapsed time ITERATION :           9.5514340   \\
== model302/helios\_b ( intelmpi + FUNNELED )                     \\
  0              \# Elapsed time ITERATION :         260.5344900   \\
  0              \# Elapsed time ITERATION :         167.9943480   \\
  0              \# Elapsed time ITERATION :           9.0590700   \\
  0              \# Elapsed time ITERATION :           9.5267600   \\
  0              \# Elapsed time ITERATION :           9.0850900   
\end{alltt}}}}\\[2mm]
From the first two commands, one can draw two conclusions:\\[-6mm]
\begin{enumerate}
\itemsep=0pt\topsep=0pt\partopsep=0pt%
\parskip=0pt\parsep=0pt%
\item The solve step (with the \tit{funneled} version of Pastix) takes roughly
   the same amount of time with \tet{bullxmpi} or \tet{intelmpi} MPI libraries.
\item The factorisation is much quicker with \tet{bullxmpi} implementation.
\end{enumerate}
The last command gives the time spent for some temporal iterations. It
is clear that \tet{bullxmpi} fastens JOREK a lot compared to
\tet{intelmpi}. The first two iterations are dominated by the
factorisation step and the \tet{intelmpi} is much slower on this
step.

\newpage
\subsubsection{Impact of the Fortran compiler}

JOREK numerical core relies partly on PASTIX and BLAS libraries. These
libraries should be configured and compiled properly. But, the
compilation of JOREK impacts also the performance. We want to evaluate
the impact of fortran compiler on timers.\\[2mm]
\scalebox{.80}{\colorbox{lightgray}{
\parbox{1.05\linewidth}{
\begin{alltt}
\$ benchmark/timing\_bench.sh ITER out\_loop2 grep5  rheticus\_gf\_302 if\_302        \\
== rheticus\_gf\_302 ( gfortran 4.4.6 )                                            \\
  0              \# Elapsed time ITERATION :         157.0370000            \\
  0              \# Elapsed time ITERATION :          78.8670000            \\
  0              \# Elapsed time ITERATION :          14.6350000            \\
  0              \# Elapsed time ITERATION :          14.9580000            \\
  0              \# Elapsed time ITERATION :          15.1500000            \\
== rheticus\_if\_302 ( ifort 12.1 )                                                    \\
  0              \# Elapsed time ITERATION :         150.9672940            \\
  0              \# Elapsed time ITERATION :          74.2307940            \\
  0              \# Elapsed time ITERATION :           8.2149000            \\
  0              \# Elapsed time ITERATION :           9.0045110            \\
  0              \# Elapsed time ITERATION :           9.8495460            
\end{alltt}}}}\\[2mm]

Each iteration takes a bit more time with \tet{gfortran} compared to
\tet{ifort}. After some investigations in the output files, the
responsible for this difference can be found:

\scalebox{.80}{\colorbox{lightgray}{
\parbox{1.05\linewidth}{
\begin{alltt}
\$ benchmark/timing\_bench.sh construct\_ out\_loop2 grep5  rheticus\_gf\_302 rheticus\_if\_302 \\
== rheticus\_gf\_302 ( gfortran 4.4.6 )                                                \\
  0       \# Elapsed time construct\_matrix :          11.9530000           \\
  0       \# Elapsed time construct\_matrix :          11.5190000           \\
  0       \# Elapsed time construct\_matrix :          11.4940000           \\
  0       \# Elapsed time construct\_matrix :          11.4600000           \\
  0       \# Elapsed time construct\_matrix :          11.5520000           \\
== rheticus\_if\_302 ( ifort 12.1 )                                                   \\
  0       \# Elapsed time construct\_matrix :           5.2446580           \\
  0       \# Elapsed time construct\_matrix :           4.8440020           \\
  0       \# Elapsed time construct\_matrix :           4.8604400           \\
  0       \# Elapsed time construct\_matrix :           4.8635080           \\
  0       \# Elapsed time construct\_matrix :           5.4503650           
\end{alltt}}}}\\[2mm]

The matrix construct subroutine takes more time with
\tet{gfortran}. It explains the gap between elapsed time for a
complete ITERATION seen previously.

\subsubsection{PASTIX thread-funneled versus thread-multiple}

The PASTIX sparse parallel solver uses by default the
MPI\_THREAD\_MULTIPLE if the MPI library supports it. The design of
this parallel library is based on a fully multi-threaded environment,
and the MPI\_THREAD\_MULTIPLE mode of MPI helps for that.  But if the
MPI implementation does not handle MPI\_THREAD\_MULTIPLE, a funneled
version of PASTIX is also accessible. This version, available through
-DPASTIX\_FUNNELED during PASTIX configuration requires only the
MPI\_THREAD\_FUNNELED capability of the MPI library. In this mode, the
MPI process may be multi-threaded, but \tit{only the main thread} will
make MPI calls.\\[2mm]
\scalebox{.80}{\colorbox{lightgray}{
\parbox{1.15\linewidth}{
\begin{alltt}
\$ cd benchmark\\
\$ ./timing\_bench.sh facto out\_loop3 grep2 model302/fourmi\_? model302/rheticus\_?\\
== model302/fourmi\_a ( openmpi )                                 \\
  0                \#\# Elapsed time, facto :          64.7742300 \\
== model302/fourmi\_b ( openmpi + FUNNELED )                      \\
  0                \#\# Elapsed time, facto :          50.9117520 \\
== model302/rheticus\_a ( mvapich2 )                              \\
  0                \#\# Elapsed time, facto :          39.4410000 \\
== model302/rheticus\_b ( mvapich2 + FUNNELED )                   \\
  0                \#\# Elapsed time, facto :          35.4110000 
\end{alltt}
}}} \\[2mm]
This example show the time taken by the factorisation step performed by
PASTIX solver on two machines for model \tet{302}.
On the two machines, with two different MPI libraries, the \tit{funneled}
version of Pastix is quicker than the \tit{multi-threaded} one.\\[2mm]
\scalebox{.80}{\colorbox{lightgray}{
\parbox{1.15\linewidth}{
\begin{alltt}
\$ cd benchmark\\
\$  ./timing\_bench.sh gmres/solve out\_loop3 grep1 model302/fourmi\_? model302/rheticus\_?\\
== model302/fourmi\_a ( openmpi )                                \\
  0            \# Elapsed time gmres/solve :           1.5070440 \\
== model302/fourmi\_b ( openmpi + FUNNELED )                     \\
  0            \# Elapsed time gmres/solve :           4.4459140 \\
== model302/rheticus\_a ( mvapich2 )                             \\
  0            \# Elapsed time gmres/solve :           1.3680000 \\
== model302/rheticus\_b ( mvapich2 + FUNNELED )                  \\
  0            \# Elapsed time gmres/solve :           3.7280000 
\end{alltt}
}}}\\[2mm]
Concerning the gmres/solve iterative process that mainly involves the
solve step of the Pastix solver, the \tit{funneled} version is nearly
three times slower than the multithreaded version. The performance ratio
is the opposite of the one made for the factorisation.

As a conclusion, if you have access to a library that has both
\tet{MPI\_THREAD\_FUNNELED} and \tet{MPI\_THREAD\_MULTIPLE} available,
you have to check which configuration gives you best performance for
your runs. If your JOREK jobs are mainly dominated by \tet{solve}
performance \tet{MPI\_THREAD\_MULTIPLE} should be better, but if your
runs are dominated by \tet{facto} performance
\tet{MPI\_THREAD\_FUNNELED} should be faster.

\subsubsection{Parallel strong scaling}

The \tet{timing\_bench.sh} tool can be used for parallel benchmarking.
In order to evaluate parallel performance of a code, a classical test
is the strong scaling. Given a fixed test case (we have taken the
reference scenario of model 302 with X-point geometry and
\tet{n\_tor=3}), the number of cores is doubled successively. Ideally,
the timing of parallel subroutines should be successively divided by a
factor two.  For this experiment, we have set the JOREK compilation
flag \tet{-DUSE\_BLOCK} that shortens the analysis step of the
parallel sparse solver. The \tet{rheticus} machine of M\'{e}socentre
de Marseilles was used.


\hspace*{-1cm}\begin{minipage}{.51\linewidth}
\begin{table}[H]
\newcommand{\bgstep}[1]{\textbf{\tiny {#1}}}
\renewcommand{\baselinestretch}{1.5}
\centering
{\scriptsize
\begin{tabular}{|p{1.8cm}||c|c|c|c|c|}
\hline
         Nb cores     & 12 & 24 & 48 & 96 & 192 \\
         Nb nodes     & 1  &  2 &  4 &  8 & 16 \\
         Nb MPI proc. & 4  &  4 &  4 &  8 & 16 \\
         Nb threads   & 3  &  6 & 12 & 12 & 12 \\
\hline
\textbf{Steps}&\multicolumn{5}{c}{}\\
\hline
     \tet{construct\_m     } &  \tit{14.74} & 7.53 & 4.13 & 2.08 & 1.47\\
     \tet{coicsr           } &  \tit{0.41 } & 0.30 & 0.29 & 0.31 & 0.31 \\
     \tet{distribute       } &  \tit{3.30 } & 2.90 & 2.93 & 2.50 & 2.25\\
     \tet{analysis         } &  \tit{2.03 } & 1.69 & 1.71 & 2.16 & 2.40 \\
     \tet{facto            } &  \tit{265.8} & 67.2 & 41.5 & 22.7 & 14.5 \\
     \tet{gmres/solve      } &  \tit{18.5 } & 2.53 & 1.76 & 1.33 & 0.73 \\
\hline
\hline
     \tet{ITER}              &  \tit{308.7} & 83.4 & 53.3 & 31.9 & 22.5 \\
\hline
\end{tabular}
} 

\caption{Timing in seconds for the first iteration  
(reference scenario model 302, \tet{out\_loop3} file)} 
\label{perf}
\end{table}
\end{minipage}\hspace*{.8cm}\begin{minipage}{.51\linewidth}
\begin{table}[H]
\newcommand{\bgstep}[1]{\textbf{\tiny {#1}}}
\renewcommand{\baselinestretch}{1.5}
\centering
{\scriptsize
\begin{tabular}{|p{1.8cm}||c|c|c|c|c|}
\hline
         Nb cores     & 12 & 24 & 48 & 96 & 192 \\
         Nb nodes     & 1  &  2 &  4 &  8 & 16 \\
         Nb MPI proc. & 4  &  4 &  4 &  8 & 16 \\
         Nb threads   & 3  &  6 & 12 & 12 & 12 \\
\hline
\textbf{Steps}&\multicolumn{5}{c}{}\\
\hline
     \tet{construct\_m     } &  {15.7} & 7.20 & 3.90 & 2.07 & 1.20\\
     \tet{coicsr           } &  {0.   } & 0.   & 0.   & 0.   & 0.  \\
     \tet{distribute       } &  {\tiny 0.002} & {\tiny 0.004} & {\tiny 0.002} & {\tiny 0.002} & {\tiny 0.002}\\
     \tet{analysis         } &  {0.   } & 0.   & 0.   & 0.   & 0.  \\
     \tet{facto            } &  {0.   } & 0.   & 0.   & 0.   & 0.  \\
     \tet{gmres/solve      } &  {12.8} & 6.89 & 5.92 & 3.56 & 2.12 \\
\hline
\hline
     \tet{ITER}              &  \tit{29.9} & 14.8 & 10.2 & 5.97 & 3.58 \\
\hline
\end{tabular}
} 

\caption{Timing in seconds for the third iteration  
(reference scenario model 302, \tet{out\_loop3} file)} 
\label{perf}
\end{table}
\end{minipage}\\[3mm]

The configuration on only one node (first column of the left table) is
pathological. It is strongly penalized by a very large part of the
available memory occupied by JOREK processes (memory shortfall and
system starts swapping). That's better to exclude this configuration
in a performance analysis.

In the left table, the timers for an iteration that includes analysis
and factorisation steps is presented.  Two steps scale quite well:
\tet{construct\_matrix} and \tet{factorisation}. Three steps do not
scale at all: \tet{coicsr}, \tet{distribute}, \tet{analysis}. A step is
in the middle with some scaling but not a good one: \tet{gmres/solve}
step. For one complete iteration (\tet{ITER} keyword), the scaling is
not so bad, this is partly due to the large relative weight of the
\tet{factorisation} step.

In the right table, the timers for an iteration that does not include
analysis and factorisation steps is shown. The \tet{distribute} step
takes less time than previous table because only the RHS vector $b$ is
transferred among cores, not the whole matrix $A$. The execution time
for one iteration is dominated by the steps \tet{construct\_matrix}
and \tet{gmres/solve} and scales reasonably well.

\section {Conclusion}
A Non Regression Tool has been installed in JOREK code that allows the
user to check if its runs give the same results as compared to some
reference scenarii. Three reference scenarii have been designed for
models 199, 302 and 303. The comparison of one run against a reference
run is performed thanks to a specific metric encasulated into a shell
script. This metric is based on the numerical differences between
several energy Fourier modes growth rates during a given time interval
against one reference scenario.

The reference scenarii can also be used for benchmarking issues. We
have compared in different configurations some timers outputted by the JOREK
code. These timers give clues to determine a performant configuration
when porting the code on new systems.
\newpage
%
\bibliographystyle{alpha}
\bibliography{NRT-jorek}

\end{document}